\pdfoutput=1

\documentclass[a4paper,12pt]{article}
\usepackage{latexsym, amssymb, enumerate, amsmath,epsfig}
\usepackage[margin=1in]{geometry}
\usepackage{setspace,color}
\usepackage[titletoc]{appendix}
\usepackage{epstopdf}
\usepackage{graphicx}
\usepackage{pgfplots}
\usepackage{multirow}
\usepackage{hyperref}
\usepackage{float}
\usepackage{caption}
\usepackage{subcaption}
\usepackage{grffile}
\usepackage{amsthm}
\usepackage[metapost]{mfpic}
\usepackage{booktabs}
\usepackage{verbatim}
\usepackage[mathlines]{lineno}
\usepackage{parskip,array}
\usepackage{algorithm}
\usepackage{algorithmic}
\usepackage{url}
\usepackage{tikz}
\usepackage{xcolor,colortbl}
\usepackage{multirow}

\usetikzlibrary{arrows.meta, matrix,calc,shapes,trees, shadings}
\tikzset{%
	>={Latex[width=2mm,length=2mm]},
	base/.style = {rectangle, rounded corners, draw=black,
		minimum width=4cm, minimum height=1cm,
		text centered, font=\sffamily},
	start/.style = {base, fill=blue!30},
	root/.style  = {base, rounded corners=2pt, thin, align=center, fill=blue!30},
	startstop/.style = {base, fill=red!30},
	activityRuns/.style = {base, fill=green!30},
	process/.style = {base, minimum width=2.5cm, fill=orange!15,
		font=\ttfamily},
	check/.style = {base, diamond, fill=yellow!30},
	deposit/.style={base, rounded corners=6pt, thin,align=center, fill=green!30},
	withdraw/.style={base, rounded corners=6pt, thin,align=center,  fill=red!30},
	function/.style={base, rounded corners=6pt, thin,align=center, fill=yellow!30},
	it/.style={font={\scriptsize\itshape}}
}

\newenvironment{customlegend}[1][]{%
	\begingroup
	\csname pgfplots@init@cleared@structures\endcsname
	\pgfplotsset{#1}%
}{%
	\csname pgfplots@createlegend\endcsname
	\endgroup
}%

\def\addlegendimage{\csname pgfplots@addlegendimage\endcsname}

\pgfkeys{/pgfplots/number in legend/.style={%
		/pgfplots/legend image code/.code={%
			\node at (0.125,-0.0225){#1}; % <= changed x value
		},%
	},
}
\pgfplotsset{
	every legend to name picture/.style={west}
}

\theoremstyle{definition}

\theoremstyle{remark}
\newtheorem{rem}{Remark}

\makeatletter
\def\thickhline{%
	\noalign{\ifnum0=`}\fi\hrule \@height \thickarrayrulewidth \futurelet
	\reserved@a\@xthickhline}
\def\@xthickhline{\ifx\reserved@a\thickhline
	\vskip\doublerulesep
	\vskip-\thickarrayrulewidth
	\fi
	\ifnum0=`{\fi}}
\makeatother

\newlength{\thickarrayrulewidth}
\nolinenumbers

\begin{document}
	\nolinenumbers
	
	\title{xLumi: Payment Channel Protocol and Off-chain Payment in Blockchain Contract Systems}
	
	\author{
		Ningchen Ying\thanks{V Systems. Email: {\bf ningchen@v.systems}}, 
		Tsz-Wai Wu\thanks{V Systems. Email: {\bf tszwai@v.systems}}
	}
	
	\maketitle
	
	\begin{abstract}
		In this paper, we introduce Super Luminal ("xLumi"), a new payment channel protocol for blockchain systems. xLumi is a simple unidirectional payment channel that can be extended to a bidirectional payment channel or a complete network. xLumi guarantees the security of the payment channel's funds by using a simple set of mathematical rules that can be easily implemented on any blockchain with the necessary infrastructure. We also give the detailed implementation methods of this idea using V Systems contract systems in this paper.
	\end{abstract}
	
	\section{Introduction}
	\label{Intro}
	Since the introduction of blockchain technology, first used by the Bitcoin protocol and subsequently by a large number of protocols \cite{nakamoto2008bitcoin, wood2014ethereum}, scalability has been a popular topic of discussion \cite{8539529, croman2016scaling}. Blockchain transactions typically are associated with high costs, and their maximum transaction throughput is too low to use as the main payment method globally. Reaching consensus requires every participating node in the blockchain to store the entire transaction history, making fees necessary to prevent spam.  Larger block sizes offer higher throughput but do not alleviate the problem of wasted storage space while introducing further security and sustainability concerns.
	
	One proposed solution to this problem is payment channels: a second layer solution to the blockchain scalability issue. Second layer protocols augment blockchain protocols without changing anything in the underlying blockchain. The payment channels attempt to take the majority of transactions off-chain and settle the final state with a single on-chain transaction. Only the two involved parties of any payments are required to know the details of their transactions, so it is unnecessary for the entire blockchain to know every single transaction that happened between them. It is possible for  transacting parties to continue to pay each other, delaying the final settlement to a single on-chain transaction when they are satisfied with the final state. This can drastically reduce the money and time needed to perform all the transactions, resulting in lower on-chain stresses.
	
	\subsection{Lightning Network}
	One of the more well known applications that resulted from payment channels are payment channel networks such as the Lightning Network \cite{poon2016bitcoin, prihodko2016flare}. The Lightning Network allows all parties to use other opened payment channels as hubs of financial capacity to pay anyone else on the network reachable through the channels the sender's node is connected to. Lightning reduces the need to open a payment channel with every party on the network you plan to transact with, further reducing the number of on-chain transactions and reducing the cost and time required for transactions. It turns payment channels from involving only two parties into a method of doing transactions with anyone within a network.  If the Lightning Network expands sufficiently, it can potentially become the main method of paying people in any blockchain protocol.
	
	Despite the Lightning Network whitepaper being very clear in design of how it will utilise payment channels in theory, the Lightning network appears to come with significant disadvantages that hinder it from being highly adopted \cite{jamesonLightning}. The Lightning Network's most obvious disadvantage is that everyone along your path to the party you are transacting with must be online and behave as expected when you decide to make your transaction, otherwise you would have to find a different path. The method of routing and implementing fees are discussed extensively \cite{prihodko2016flare, di2018routing}.
	
	This could potentially be solved by several large organisations providing stable central hubs that have payment channels with a large number of users, this opens up to the protocol being significantly more centralised than it would like to be. The network also appears to be vulnerable to flood attacks, whereby attackers flood a large amount of channels by generating lots of transactions at the same time. This may cause the network to be unable to confirm some claiming transactions. If the attack is sustained for long enough, the claiming transactions may expire, causing the claimant to lose their funds, and the attacker will be able to obtain the rest of the funds in the payment channel. \cite{harris2020flood}.
	
	\subsection{Sidechains}
	Sidechains are another proposed method of solving scalability issues in blockchain protocols \cite{back2014enabling}. Due to the nature of how changes to blockchain protocols are made, it is very difficult to change any part of it. It requires a very high threshold of community consensus, and often come with some backlash from users. Sidechains seek to solve this problem through a two-way peg between two blockchains, allowing users to transfer main chain coins for coins on this sidechain and vice versa, usually at a set exchange rate. This means that it would be possible for main chain users to utilise the advantages of the sidechain without the need to change the main chain's protocol. Of course, it is not possible to truly transfer main chain coins to the sidechain, so the current method adopted is to send the coins to a main chain address that essentially locks up the coins and issues the corresponding number of coins on the other chain.
	
	This usually comes in the form of a multi-gateway releasing a coin on the sidechain once a coin has been sent to it, and vice versa. This is useful in that transactions can be performed on the sidechain as if they had currency from the main chain. Since the sidechain may provide some useful services, such as allowing more powerful smart contracts or faster transaction speeds, sidechains should store all of the states of its parent chain and ideally it would be no different to having the main chain's functionality extended. In the case of DEX's (decentralised exchanges), a similar technique of locking and releasing coins on different chains is used, however, DEX's are simply separate blockchains that utilises two-way pegs with many chains to allow trade between them, and does not store their states.
	
	RSK  \cite{lerner2015rsk} is a Bitcoin sidechain, aiming to enable turing complete smart contracts, and faster transaction confirmations. There are security concerns for sidechains due to a difference in computing power between the parent chain and its child chains. Child chains typically have significantly less computing power to maintain consensus, opening up the possibility for miners in the parent chain to attack child chains. RSK currently solves these security issues, including the job of locking up and releasing coins on each blockchain, by utilising a trusted third party. This is of course, not ideal, as a third party introduces an extra point of weakness in the protocol.
	
	\subsection{Statechains}
	Another recent attempt in this space are Statechains, which introduces a trusted third party into payment channels \cite{somsen2018statechains}, taking away a number of disadvantages of payment channels. In Statechains, trusted third parties ensure that the final state of the payment channels are correct and therefore, in theory it should be impossible to cheat as long as these trusted third parties act honestly. The protocol solves the routing and capacity problems that can be seen in the Lightning Network, as the routes are dealt with by a mediator. The disadvantage of Statechains is that they introduce a third party which can potentially steal by colluding with previous holders of the coins. However, collusion is provable in the Statechain design and if it happens, the third party will lose their trusted status in the network. Other users going through the dishonest third party can immediately close their payment channels on the chain and minimize their losses. A second unique disadvantage of Statechains is that the proposed method of transferring funds involve handing over private keys of specific UTXOs. This means that the protocol can only support the transfer of entire UTXOs (Unspent Transaction Outputs), which can be a real issue when trying to send over specific amounts of tokens, since parties may not have UTXOs of that value.
	
	\subsection{Scaling Blockchain}
	There have been many ideas and implementations over the years to improve blockchain protocols, extending their usages and improving transaction processing capacity. Beyond what we have already discussed, blockchain as a technology has made tremendous progress throughout the years. The number of breakthroughs are vast and it would be impossible to discuss them all within a single Whitepaper, however, a few other noteworthy protocols that inspired the creation of xLumi are, a protocol similar to Bitcoin's Lightning Network called the Raiden Network \cite{raidenNetwork} built on Ethereum, or the Duplex Micropayment Channel network \cite{decker2015fast} which utilises pairs of unidirectional payment channels to create a network.
	
	The biggest advantage of the aforementioned methods is that they do not require any changes to the protocol. They solve the scalability issues by taking transactions away from the main chain, and settle them in some other fashion, such as locally or on a sidechain, although they often introduce a number of new security concerns.
	
	There appears to often be a trade off between security and scalability. Bitcoin provides a good example of this trade off; its core consensus mechanism, Proof of Work (POW), is considered extremely robust and the 51\% attack being deemed infeasible. Proof of Work's current limitation is that it uses a large amount of computing power and storage space, and is therefore extremely difficult to scale. There are many areas where increasing a blockchain protocol's security reduces its scalability.  Although a lot of security issues can be circumvented without sacrificing scalability and performance by employing a trusted third party. While the use of trusted third parties is potentially a good trade off in certain situations, there needs to be a lot more considerations when using them, in particular, the result of dishonest behaviour should be known, and losses must be kept within reason.
	
	To address some of these issues, Super Luminal ("xLumi"), a new protocol for creating payment channels is introduced in this paper. xLumi can be used by any blockchain protocols that can store simple states values, therefore, most blockchains that allow smart contracts on their platform should be able to use xLumi. This protocol will be a unidirectional payment channel, such that the funds can only flow in one direction. The stored states will ensure the security of the funds and uniqueness of the payment channel states.
	
	While xLumi has its own advantages and disadvantages with regards to scalability, there is one extra barrier that blockchain communities must consider when introducing protocols: {\bf How easy is it to implement and use?} Some key ideas that drives xLumi's design is clarity of design and ease of use for users. xLumi's simplicity allows any blockchain with suitable functionality to implement it. While current implementations of xLumi is very simple, it can be easily expanded to allow for more complex use cases.
	
	We will be going through some necessary concepts to understand the design of xLumi in the Section \ref{sec:concepts}, and go into more detailed explanations in Section \ref{sec:xLumi}. The current implementation of the new unidirectional payment channel is described in Section \ref{sec:impl}, and some possible extensions to the protocol are discussed in Section \ref{sec:ext}. We will give some usage cases in Section \ref{sec:uses}, and conclude our protocol in Section \ref{Summary}.
	
	\section{Basic Concepts}
	\label{sec:concepts}
	In the following section, necessary concepts and their features are described in detail. The first part deals with the fundamental properties of blockchain digital signatures, and readers who are familiar with the topic of digital signatures should be free to skip Section \ref{sec:digitalSig}. Section \ref{sec:motivations} gives an insight into our motivations for developing a new payment channel protocol, hopefully making the design decisions of the protocol clearer for the reader.
	
	\subsection{Digital Signatures and Its Usage in Blockchain Systems}
	\label{sec:digitalSig}
	Digital signature schemes are mathematical schemes which are used to verify the authenticity of digital messages or documents. A valid signature gives the recipients a strong assurance that the message or document was agreed on by a \textbf{known} sender. At the same time, it also helps to verify that the message was not changed in transit. These give the digital signature two important properties: \textbf{authenticity} and \textbf{integrity}. With these reasons, digital signature schemes have constructed the digital world in last several decades, and it is also fundamental to blockchain systems..
	
	In general, digital signature comes in two parts, the message and the signature.
	\begin{equation}
	(M, Sign(M))
	\end{equation}
	The exact method of signing $M$ is what distinguishes particular signature schemes from each other, each having their own strengths and weaknesses.
	
	Some prerequisites are required to get digital signatures. A key generation algorithm is needed to get a secret key (which is kept private, and is used to sign messages) and public key (known by the recipients/public to verify the messages).
	
	Some crucial properties in the digital signatures that are used in blockchain systems are:
	\begin{center}
		\begin{enumerate}
			\item Verifiable sender, signatures generated by a particular sender must be distinguishable from any other senders except with negligible probability.
			\item 
			\label{uniqueSig}
			Unique signatures, distinct messages must generate distinct signatures to ensure messages are distinguishable except with negligible probability.
			\item Authenticity of the message, it should be impossible for anyone other than the sender to change the message and still have a valid signature except with negligible probability.
		\end{enumerate}
	\end{center}
	
	Several well established cryptographic signature schemes have these properties proven through extensive usage. Including DSA \cite{schneier2007applied} which relies on the difficulty of the discrete logarithm problem, Bitcoin's ECDSA \cite{johnson2001elliptic}, and other signature schemes that use the Elliptic Curve discrete logarithm problem such as EdDSA \cite{josefsson2017edwards}.
	
	\begin{rem}
		Signing the same message with different secret keys will get distinguishable signatures. Different messages signed by the same secret key will get distinguishable signatures too. In some cryptographic signature schemes, the same message signed by the same secret key will also get distinguishable signatures by using a random nonce.
	\end{rem}

	\begin{rem}		
		One important feature in most blockchain systems is that messages signed by individual users can only be used as a valid information once. For instance, the same transaction from the sender to the recipient can only be recorded as valid transaction once. Digital signatures makes ensuring this property extremely simple and is currently the most widely chosen method to represent blockchain transactions.
	\end{rem}

	\subsection{Payment Channel Usages and Challenges}
	\label{sec:motivations}
	It is still debated by blockchain communities whether it is essential to increase the maximum throughput of blockchains by speeding up the transaction verification, expanding the block size to support more transactions in one block or some other change in the underlying protocol. Some of these debates have ended in hard forks of the protocols, the most noteworthy being Bitcoin forks such as Bitcoin Cash or Bitcoin SV. One of the critical debate points is the method to judge the type of information that should be recorded in the blockchain database. There are several on-chain solutions to control or improve the blockchain database, and some of which have been implemented in many projects, but the effectiveness of these solutions remains to be seen. Another direction of solutions is using off-chain ideas in order to keep some information offline. These solutions will be helpful in on-chain database storage problems, privacy and on-chain network load problems, but it can also weaken the security of blockchain transactions when compared to recording everything on-chain.
	
	Payment channel protocols is one of these off-chain solutions to improve user experience when using the blockchain. The key idea of most payment channel protocols is to keep only important states on-chain while high-frequency interactions remain offline. The interaction between the on-chain and off-chain states give developers a large amount of room to draw from imagination and several excellent ideas were born to solve the challenges blockchain protocols face. One popular implementation of payment channels is the punishment payment channel, which is one of the first proposed payment channel protocols. This method is also what is used in the Bitcoin Lightning Network.
	
\begin{figure}[H]
		\scriptsize
		\centering
		
		% Drawing part, node distance is 1.7 cm and every node
		% is prefilled with white background
		\begin{tikzpicture}[node distance=1.7cm,
		every node/.style={fill=white, font=\sffamily}, align=center]
		% Specification of nodes (position, etc.)
		\node (start)             [start]              {Create Channel};
		\node (createWalletBlock)     [process, below of=start]          
		{Create a 2-of-2 multi-signature wallet};
		\node (refundBlock)      [process, below of=createWalletBlock]   
		{Obtain a refund transaction};
		\node (fundBlock)     [process, below of=refundBlock]   
		{Fund the multi-signature wallet};
		\node (useofflineBlock)      [activityRuns, below of=fundBlock]
		{Use the payment channel offline};
		\node (offLineTxBlock)    [start, right of=start, xshift=5.5cm]
		{Offline Transactions};
		\node (transitoryKeyBlock) [process, below of=offLineTxBlock, text width = 4cm]
		{Create a transitory key
			(some generated secret)};
		\node (signTxBlock) [process, below of=transitoryKeyBlock, text width=4cm]
		{Sign a payment transaction};
		\node (passSignatureBlock) [process, below of=signTxBlock, text width=4cm]
		{Pass valid signature to other party};
		\node (exchangeTransitoryKeyBlock) [activityRuns, below of=passSignatureBlock, text width=4cm]
		{Exchange tansitory keys of old states};
		\node (close) [start, below of=exchangeTransitoryKeyBlock, text width=4cm, xshift=-5cm]
		{Closing the channel};
		\node (broadcastBlock) [startstop, below of=close, text width=4cm, xshift=-2cm, yshift=-1cm]
		{Broadcast the latest payment transaction};
		\node (checkBlock) [check, below of=close, text width=2cm, xshift=3cm, yshift=-2.5cm]
		{Check if state is correct};
		\node (correctStateBlock) [startstop, below of=checkBlock, text width=2cm, xshift=-2.3cm, yshift=-2cm]
		{End};
		\node (incorrectStateBlock) [startstop, below of=checkBlock, text width=4cm, xshift=3cm, yshift=-2cm]
		{Broadcast same transaction with transitory key};
		% Specification of lines between nodes specified above
		% with aditional nodes for description 
		\draw[->]             (start) -- (createWalletBlock);
		\draw[->]     (createWalletBlock) -- (refundBlock);
		\draw[->]      (refundBlock) -- (fundBlock);
		\draw[->]     (fundBlock) -- (useofflineBlock);
		\draw[->]     (offLineTxBlock) -- (transitoryKeyBlock);
		\draw[->]      (transitoryKeyBlock) -- (signTxBlock);
		\draw[->]     (signTxBlock) -- (passSignatureBlock);
		\draw[->]     (passSignatureBlock) -- (exchangeTransitoryKeyBlock);
		\draw[->]    (close) -- node[xshift=-1cm]{Broadcasting party}(broadcastBlock);
		\draw[->]     (close) -- node[xshift=1cm]{Non-broadcasting party}(checkBlock);
		\draw[->]    (checkBlock) -- node{Yes}(correctStateBlock);
		\draw[->]     (checkBlock) -- node{No}(incorrectStateBlock);
		\end{tikzpicture}
		\caption{Flow diagram of the punishment payment channel. The protocol including three parts: create channel, offline transactions and closing the channel. Broadcasting the transaction with the transitory key allows the party to obtain all the funds in the channel. \label{fig:punishmentFlow}}
	\end{figure}
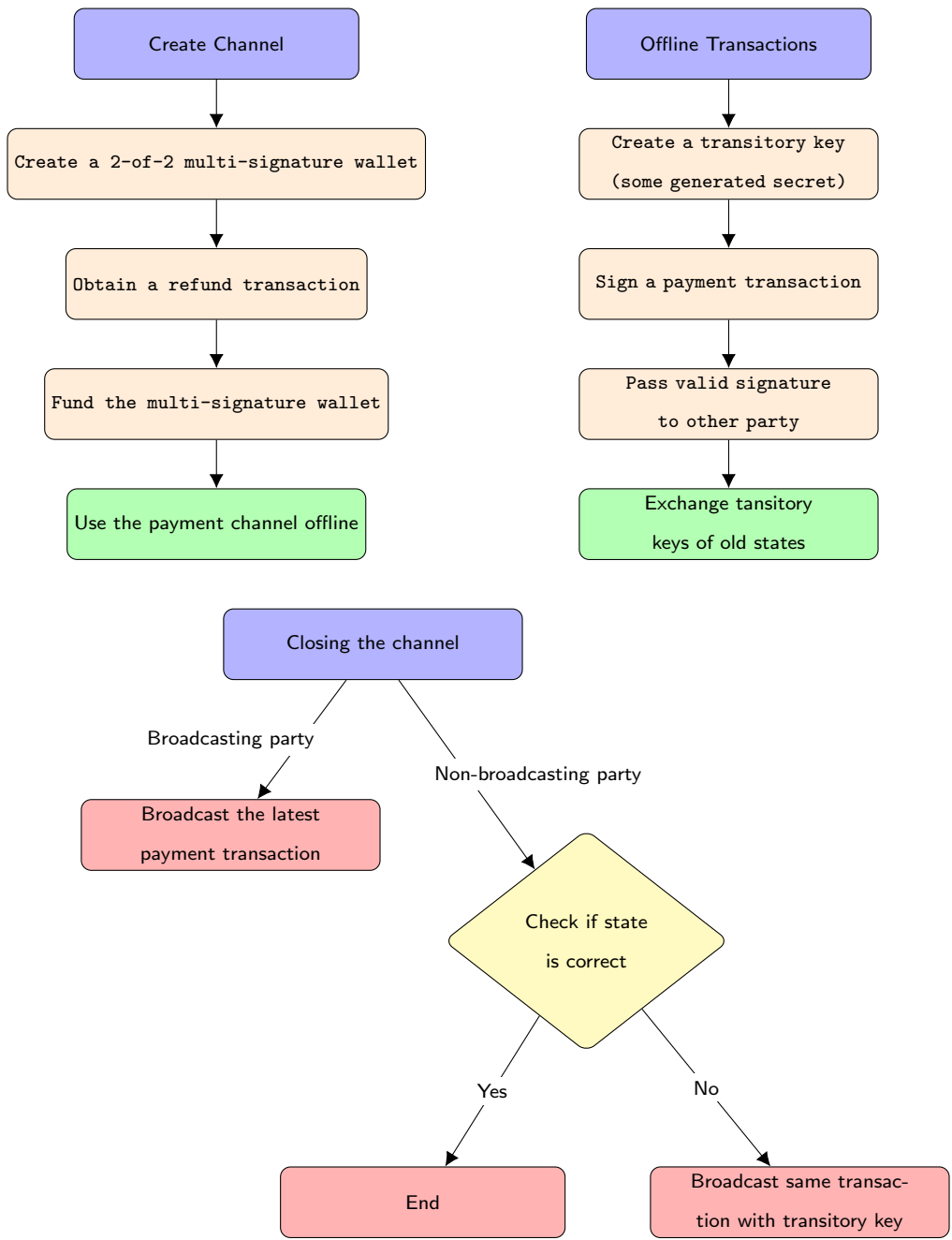

	Using the Lighting Network payment channel as an example, the basic idea for this payment channel is that by using a 2-party multi-signature wallet, either party can sign transactions offline that output different amounts to the two parties. The correct state of the payment channel is ensured by punishing parties that attempt to broadcast old states. If one of the two parties attempts to cheat by broadcasting an older state, the protocol allows the other party to obtain all the funds in the channel, the details of which are described in \cite{poon2016bitcoin}. The details of the procedures to use a punishment payment channel is presented in Figure \ref{fig:punishmentFlow}.
	
	\begin{rem}
		There are several methods that can manage multi-owner assets. One basic idea is the multi-signature or threshold signature, another commonly used idea in blockchain systems with contracts is to use contract states with extra conditional controls.
	\end{rem}
	
	The Punishment payment channel is a useful example in the design of the on-chain and off-chain interaction in payment channel protocols, but it also presents many challenges for users. For example, long-standing payment channels require a large amount of storage space as the number of transactions build up. Since it is necessary to store the transitory key of every transaction that isn't the latest one. However, it has been suggested that it is possible to use certain tricks to reduce the number of keys that users need to store, such as ensuring the transitory keys form a hash tree such that keys lower on the tree can be computed instead of needing to be stored \cite{dryja2018scalingbitcoin}.
	
	It is also argued that the cheating party may not intentionally have done so, software bugs or accidents may cause the party to lose the latest transaction, which may result in either, the payment channel users being unable to close the channel, or to close the channels with an old state. Since using an old state to close the channel could cause them to lose their funds, this is not a viable option.
	
	The next issue is that the level of punishment is not the same across payment channels, and depends entirely on how much funds were in the channel in the first place, and how much of those funds belongs to each party in the final state.  A user's incentive to cheat depends entirely on the evaluation between the risk and reward, it is possible for the state of the channel to be in a position where the risk for one party is very limited while cheating successfully is very rewarding.
	
	The large number of interactions at every transaction necessary to ensure the security of the funds in the payment channel is also not ideal. Each interaction creates more places where faults can appear.
	
	These challenges and issues give us the motivation to design a new protocol that will be helpful in solving some of the problems or improve the user experience when compared to the punishment payment channel protocol. In the next section, we will provide the details of the design in this protocol.
	
	\section{xLumi: a Unidirectional Payment Channel Protocol}
	\label{sec:xLumi}
	Having introduced the punishment payment channel, it should be noted that it can be used as either a unidirectional payment channel or a bidirectional payment channel. Using similar ideas, the unidirectional punishment payment channel would only require the payer to be punished, since newer states will always have a larger payment amount to the recipient, broadcasting old states can only benefit the payer.
	
	Unidirectional payment channels can be achieved using a number of different ideas, in this paper, a unidirectional payment channel protocol named xLumi is introduced that utilises a set of simple mathematical constraints to ensure its security.

	\subsection{Details of xLumi}
	Compared to current implementations of punishment payment channels on Lighting Network, xLumi ensures the correct state of the payment channel without the need to punish malicious parties. By using smart contracts to store state variables and control the funds, the correct state can be ensured. This drastically reduces the complexity of payment channels and also reduces the number of interactions and storage of keys required at every transaction.
	
	An xLumi channel stores two on-chain values that ensure the security of the funds in the payment channel. There is also one extra important value hidden from the blockchain, the accumulated payment amount. If the implementation requires it, another state value of expiration timestamp can be added to allow the channel to expire, this is not strictly necessary as an indefinitely valid payment channel can be useful in certain scenarios.
	
	\newpage
	We define three variables that are used in this protocol:
	\begin{itemize}
		\item $X$: Accumulated Load
		\item $Y$: Accumulated Collect
		\item $Z$: Accumulated Payment (off-chain)
	\end{itemize}
	
	These variables all form step functions with respect to time (each transaction represents a step), and it is important to note that they are all increasing monotone functions. We ensure that $X \geq Z \geq Y$. $X$, $Y$ and $Z$ may have several intersections. The receiver can collect a maximum of the amount the payer has deposited into the channel, but can only collect as much as the payer has given them the signature for. The payer is able to increase the amount they deposit into the channel, and signed transactions above the current amount deposited should be ignored by the recipient.

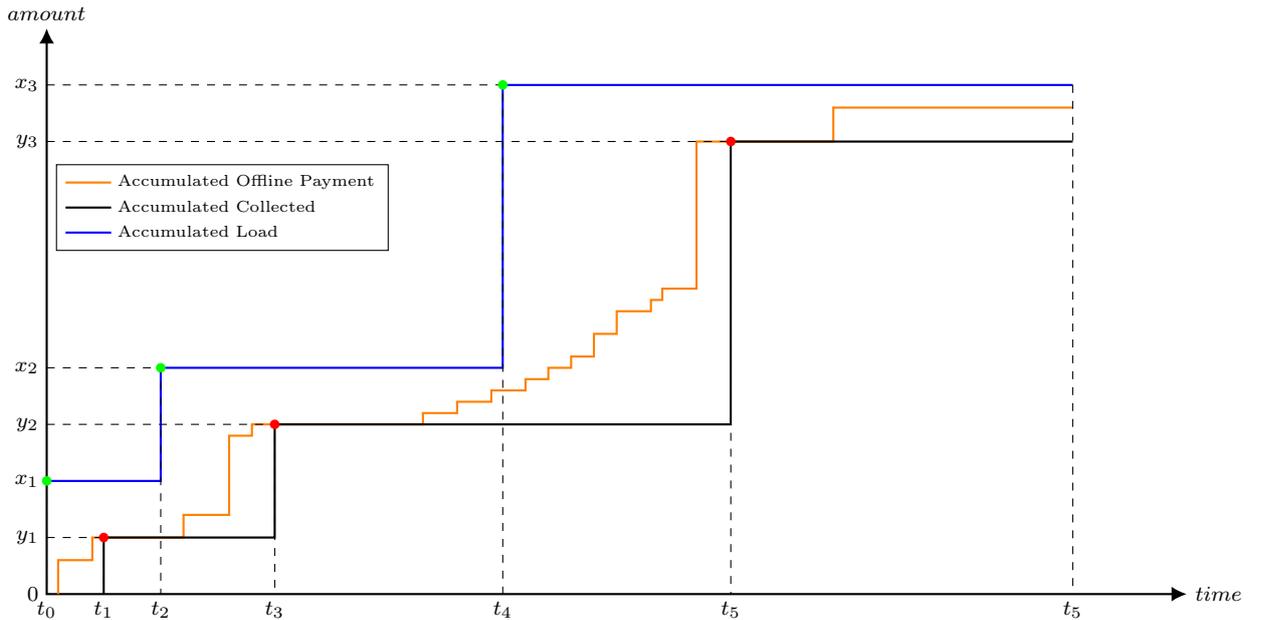
\begin{figure}[H]
	\centering
	\scriptsize
	\begin{tikzpicture}[scale=1.5]
	% Draw axes
	\draw [<->,thick] (0,5) node (yaxis) [above] {$amount$}
	|- (10,0) node (xaxis) [right] {$time$};
	% Draw X value step function
	\draw[blue, thick] (0,1) -| (1,2) -| (4,2) -| (4,4.5) -- (9,4.5);

	% Draw Z value step function
	\draw[orange, thick] (0.1,0) |- (0.4,0.3) |- (1.2,0.5) |- (1.6,0.7) |- (1.8,1.4)
	|- (3.3,1.5) |- (3.6,1.6) |- (3.9, 1.7) |- (4.2,1.8) |- (4.4,1.9) |- (4.6, 2) |- (4.8, 2.1) |- (5, 2.3) |- (5.3,2.5) |- (5.4, 2.6) |-  (5.7,2.7)
	|- (6.9,4) |- (9,4.3);
	
	% Draw Y value step function
	\draw [black, thick] (0.5,0) |- (2,0.5) |- (2,1.5) |- (6,1.5) |- (6,4) |- (9,4);
	
	% Calculate the intersection of the lines a_1 -- a_2 and b_1 -- b_2
	% and store the coordinate in c.
	\coordinate (d_1) at (0.5, 0.5);
	\coordinate (d_2) at (2, 1.5);
	\coordinate (d_3) at (6, 4);
	\coordinate (a_1) at (0, 1);
	\coordinate (a_2) at (1, 2);
	\coordinate (a_3) at (4, 4.5);
	% Draw lines indicating intersection with y and x axis. Here we use
	% the perpendicular coordinate system
	\draw[dashed] (yaxis |- d_1) node[left] {$y_1$}
	-| (xaxis -| d_1) node[below] {$t_1$};
	\draw[dashed] (yaxis |- d_2) node[left] {$y_2$}
	-| (xaxis -| d_2) node[below] {$t_3$};
	\draw[dashed] (yaxis |- d_3) node[left] {$y_3$}
	-| (xaxis -| d_3) node[below] {$t_5$};
	
	\draw[dashed] (yaxis |- a_1) node[left] {$x_1$}
	-| (xaxis -| a_1) node[below] {$t_0$};
	\draw[dashed] (yaxis |- a_2) node[left] {$x_2$}
	-| (xaxis -| a_2) node[below] {$t_2$};
	\draw[dashed] (yaxis |- a_3) node[left] {$x_3$}
	-| (xaxis -| a_3) node[below] {$t_4$};
	\draw[dashed] (9,4.5) -- (9,0) node[below] {$t_5$};
	
	\draw (0,0) node[left] {$0$};
	
	% Draw a dot to indicate intersection point
	\fill[red] (d_1) circle (1.2pt);
	\fill[red] (d_2) circle (1.2pt);
	\fill[red] (d_3) circle (1.2pt);
	\fill[green] (a_1) circle (1.2pt);
	\fill[green] (a_2) circle (1.2pt);
	\fill[green] (a_3) circle (1.2pt);
	\begin{customlegend}[legend cell align=left, %<= to align cells
	legend entries={ % <= in the following there are the entries
		Accumulated Offline Payment,
		Accumulated Collected,
		Accumulated Load
	},
	legend style={at={(3,3.8)},font=\tiny}] % <= to define position and font legend
	% the following are the "images" and numbers in the legend
	\addlegendimage{draw = orange, thick}
	\addlegendimage{draw = black, thick}
	\addlegendimage{draw = blue, thick}
	\end{customlegend}
	\end{tikzpicture}
	\caption{Evolution of xLumi's variables over time. Each step represents a transaction. The blue line represents the accumulated load $X$, the orange line represents the accumulated offline payment $Z$, and the black line represents the accumulated collected $Y$. The green points represents the on-chain transactions by the payer. The red points represent the on-chain transactions by the recipient.\label{xlumiFig}}
\end{figure}
	The key features of xLumi can be seen here, while every step in the variables shows a transaction, the amount sent (orange) happens offline, which means repeated payments can be done without paying blockchain transaction fees. Depositing and withdrawing from an xLumi contract are settled on-chain, the transactions shown in Figure \ref{xlumiFig}, shows the number of on-chain transactions are greatly reduced. The special transactions labelled at $t_1$, $t_3$ and $t_5$ represents collect transactions by the recipient, this is where the recipient obtains the funds sent to them offline.
	
	Each offline payment transaction happens by the payer signing a payment message and passing it to the recipient. Collecting payments involve the recipient broadcasting the payer's signature to the blockchain, despite the signature being broadcast by the recipient, the properties of digital signatures which are described in \ref{sec:digitalSig}, allow both the recipient and the blockchain to be sure that the payment transaction was indeed intended by the payer.

	The number of coloured circles represent the the total number of transactions that are recorded on-chain. While the number of offline payments has no limits during the time which the payment channel persists. Once the payer signs a new payment, any old signatures can be deleted, regardless of how many transactions the payment channel is utilised for, the essential information does not exceed a single signature.
	
	There must also be a method of distinguishing between payment channels and their states. It is possible to use a variety of state variables for this purpose, as long as it is guaranteed to be unique. For example, the transaction to open a contract may have unique transaction IDs, or unique indexes.
	
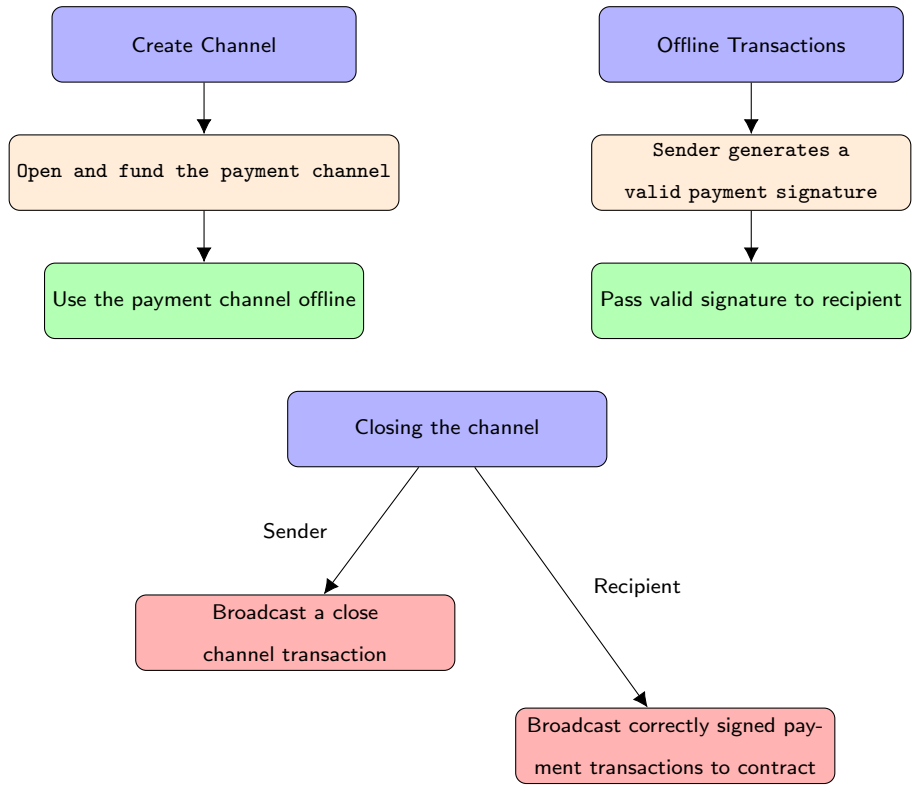
\begin{figure}[H]
		\scriptsize
		\centering
		% Drawing part, node distance is 1.7 cm and every node
		% is prefilled with white background
		\begin{tikzpicture}[node distance=1.7cm,
		every node/.style={fill=white, font=\sffamily}, align=center]
		% Specification of nodes (position, etc.)
		\node (start)             [start]              {Create Channel};
		\node (refundBlock)      [process, below of=start]   
		{Open and fund the payment channel};
		\node (useofflineBlock)      [activityRuns, below of=refundBlock]
		{Use the payment channel offline};
		\node (offLineTxBlock)    [start, right of=start, xshift=5.5cm]
		{Offline Transactions};
		\node (transitoryKeyBlock) [process, below of=offLineTxBlock, text width = 4cm]
		{Sender generates a valid payment signature};
		\node (exchangeTransitoryKeyBlock) [activityRuns, below of=transitoryKeyBlock, text width=4cm]
		{Pass valid signature to recipient};
		\node (close) [start, below of=exchangeTransitoryKeyBlock, text width=4cm, xshift=-4cm]
		{Closing the channel};
		\node (broadcastBlock) [startstop, below of=close, text width=4cm, xshift=-2cm, yshift=-1cm]
		{Broadcast a close channel transaction};
		\node (checkBlock) [startstop, below of=close, text width=4cm, xshift=3cm, yshift=-2.5cm]
		{Broadcast correctly signed payment transactions to contract};
		% Specification of lines between nodes specified above
		% with aditional nodes for description 
		\draw[->]             (start) -- (refundBlock);
		\draw[->]      (refundBlock) -- (useofflineBlock);
		\draw[->]     (offLineTxBlock) -- (transitoryKeyBlock);
		\draw[->]      (transitoryKeyBlock) --  (exchangeTransitoryKeyBlock);
		\draw[->]    (close) -- node[xshift=-1cm]{Sender}(broadcastBlock);
		\draw[->]     (close) -- node[xshift=1cm]{Recipient}(checkBlock);
		\end{tikzpicture}
		\caption[position=bottom]{Flow diagram of unidirectional payment channel protocol\label{fig:xlumiFlow}}
	\end{figure}

	The requirement that the amount paid to the recipient can only increase gives a neat way to ensure only newer states are broadcast without requiring some state index, the latest state is simply always the one with the largest amount paid. This also ensures the signature of every valid payment message is unique, since it is useless for the payer to sign a message with the same paid value for a later state. In xLumi, it is not necessary to rely on punishments or third parties to enforce final states, if the recipient party can produce a signed payment message with a higher paid value, then they can always change the state.
		
	xLumi reduces the problem of settling new states and solving the requirement of uniqueness to a simple set of mathematical rules the payment channel must follow, and does not depend on interactions between the parties utilising the channel. The Figure \ref{fig:xlumiFlow} shows the details of steps needed to utilise an xLumi payment channel.
	
	To see how funds are kept secure by simply ensuring three values do not decrease, we will first give a simple example in Section \ref{sec:xLumiExample} and the details of the implementation will be given in Section \ref{sec:impl}.

\newpage

		\subsection{Unidirectional Payment Channel Example}
	\label{sec:xLumiExample}
	We will start with opening payment channel and funding $10$ Coins into it, the initial state will be:
	
	\begin{table}[!htbp]
		\centering
		\scriptsize
		\begin{tabular}{|c|c|c|c|c|c|}\thickhline
			\multicolumn{2}{|c|}{\cellcolor[gray]{0.8}}&\multicolumn{4}{c|}{\cellcolor[gray]{0.8}\textbf{Offline States}}\\\cline{3-6}
			\multicolumn{2}{|c|}{\multirow{-2}{*}{\cellcolor[gray]{0.8}\textbf{Payment Channel States}}}& Time & Alice & Bob & On-chain\\\thickhline
			\cellcolor[gray]{0.8}Accumulated Load & $10$ Coins & $t_0$ & $0$ Coins & $10$ Coins & Yes\\\hline
			\cellcolor[gray]{0.8}Accumulated Payment & $0$ Coins &&&&\\\hline
			\cellcolor[gray]{0.8}Expiration Time & 14:00:00, May 1, 2021 &&&&\\\hline
			\cellcolor[gray]{0.8}Owner Address & Bob's Address &&&&\\\hline
			\cellcolor[gray]{0.8}Recipient Address & Alice's Address &&&&\\\thickhline
		\end{tabular}
	\end{table}

	Once the payment channel has been opened on the blockchain, Bob can begin to pay Alice offline using signed transactions. The amount of information in the transaction is very small, containing only a signature and the payment amount.
	
	\begin{equation*}
		Sign_B("\textnormal{Total paid to Alice: 1 Coin}")
	\end{equation*}
	
	The payment channel state remains the same after the offline interaction, since the blockchain does not know Bob has paid Alice. Alice now simply has a signature, which acts as a \textbf{promise} that she will be paid $1$ coin when she broadcasts the signature. The validity and contents of the signature is easily verified due to the properties of signed messages described in \ref{sec:digitalSig}.
	
	\begin{table}[!htbp]
		\centering
		\scriptsize
		\begin{tabular}{|c|c|c|c|c|c|}\thickhline
			\multicolumn{2}{|c|}{\cellcolor[gray]{0.8}}&\multicolumn{4}{c|}{\cellcolor[gray]{0.8}\textbf{Offline States}}\\\cline{3-6}
			\multicolumn{2}{|c|}{\multirow{-2}{*}{\cellcolor[gray]{0.8}\textbf{Payment Channel States}}}& Time & Alice & Bob & On-chain\\\thickhline
			\cellcolor[gray]{0.8}Accumulated Load & $10$ Coins & $t_0$ & $0$ Coins & $10$ Coins & Yes\\\hline
			\cellcolor[gray]{0.8}Accumulated Payment & $0$ Coins & $t_1$ & $1$ Coins & $9$ Coins & No\\\hline
			\cellcolor[gray]{0.8}Expiration Time & 14:00:00, May 1, 2021 &&&&\\\hline
			\cellcolor[gray]{0.8}Owner Address & Bob's Address &&&&\\\hline
			\cellcolor[gray]{0.8}Recipient Address & Alice's Address &&&&\\\thickhline
		\end{tabular}
	\end{table}
	
	It is entirely up to Alice whether she wishes to pay the transaction fee involved in updating the state of the payment channel, as long as she updates the state before the channel closes, the 1 coin is safely hers.
	
	If Bob wishes to pay Alice another Coin, he must sign another transaction.
	
	\begin{equation*}
	Sign_B("\textnormal{Total paid to Alice: 2 Coins}")
	\end{equation*}
	
	Here the two coins represent the total amount Bob wishes to pay Alice. This may be in exchange for some service or product given to Bob.
	
	Alice can choose whether or not to broadcast this transaction and pay a transaction fee. If she chooses to do so, she will update the state of the channel to:
		
	\begin{table}[!htbp]
	\centering
	\scriptsize
	\begin{tabular}{|c|c|c|c|c|c|}\thickhline
		\multicolumn{2}{|c|}{\cellcolor[gray]{0.8}}&\multicolumn{4}{c|}{\cellcolor[gray]{0.8}\textbf{Offline States}}\\\cline{3-6}
		\multicolumn{2}{|c|}{\multirow{-2}{*}{\cellcolor[gray]{0.8}\textbf{Payment Channel States}}}& Time & Alice & Bob & On-chain\\\thickhline
		\cellcolor[gray]{0.8}Accumulated Load & $10$ Coins & $t_0$ & $0$ Coins & $10$ Coins & Yes\\\hline
		\cellcolor[gray]{0.8}Accumulated Payment & $2$ Coins & $t_1$ & $1$ Coins & $9$ Coins & No\\\hline
		\cellcolor[gray]{0.8}Expiration Time & 14:00:00, May 1, 2021 &$t_2$ & $2$ Coins & $8$ Coins& Yes\\\hline
		\cellcolor[gray]{0.8}Owner Address & Bob's Address &&&&\\\hline
		\cellcolor[gray]{0.8}Recipient Address & Alice's Address &&&&\\\thickhline
	\end{tabular}
\end{table}

	Since the \textbf{Accumulated Load} of the channel can only increase, Alice will be convinced by Bob's signed transactions as long as they are for $10$ Coins or less. It is impossible for Alice to take funds from Bob that he didn't give the signature for. The funds are controlled by the payment channel, and Bob will always receive $(Accumulated Load - Accumulated Payment)$ Coins back once the channel closes.
	
	Recipients of funds from an xLumi channel can settle transactions to the chain before the channel closes without affecting the security of the funds. This may be necessary in order to minimise the opportunity cost of using the payment channel for the recipient.
	
	The next section describes the current implementation of the VSYS xLumi contract, but it is by no means the only way of implementing this type of payment channel. The implementation of the channel will depend completely on its intended use, and some possible extensions will be discussed in the Section \ref{sec:ext}

	\section{Implementation}
\label{sec:impl}
V Systems smart contracts have unique IDs based on their contract registration transaction ID that acts as special contract accounts that are able to store coins or tokens. The smart contract then has control over the coins or tokens based on its functions. Users interact with these contracts through transactions that execute the code stored on the blockchain.

The payment channel contract can be thought of as serving the function of a traditional financial institution. Funds deposited into the contract is reflected in the state of the blockchain, and deposited funds become part of the contract's balance. Within the contract, the funds can be utilized within a given set of rules. These rules are transparent and can be viewed by any party.

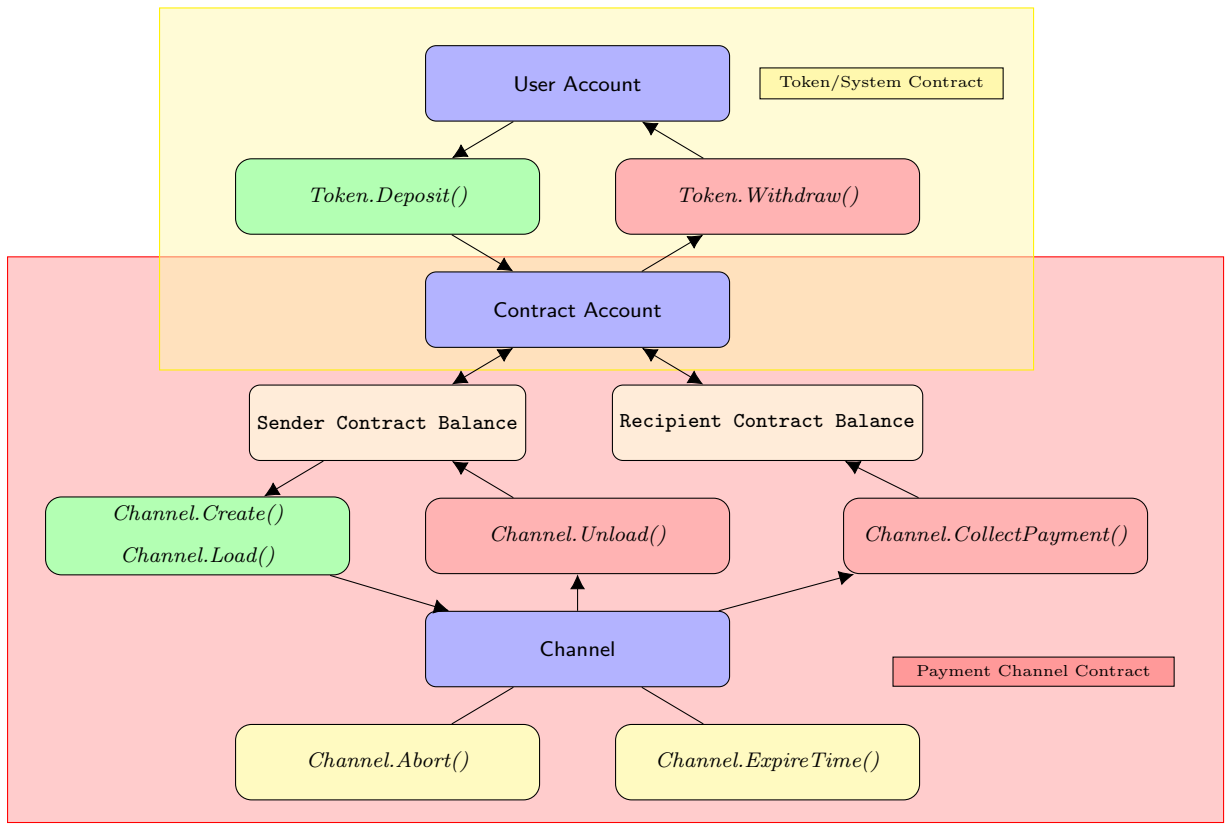
\begin{figure}[H]
	% Drawing part, node distance is 1.7 cm and every node
	% is prefilled with white background
	\scriptsize
	\centering
	\begin{tikzpicture}[node distance=1.5cm,
	level 1/.style={sibling distance=40mm},
	every node/.style={fill=white, font=\sffamily}, align=center]
	\filldraw[fill=red!40!white, draw=red, fill opacity=0.5] (8.5,-2.3) rectangle (-7.5,-9.8);
	\filldraw[fill=yellow!40!white, draw=yellow, fill opacity=0.5] (6,1) rectangle (-5.5,-3.8);
	
	\node[draw,text width=3cm, fill=yellow!40!white, font = \tiny] at (4,0) {Token/System Contract};
	\node[draw,text width=3.5cm, fill=red!40!white, font = \tiny] at (6,-7.8) {Payment Channel Contract};
	% Specification of nodes (position, etc.)
	\node (userAccount)            [start]         {User Account};
	\node (tokenDeposit)     [deposit, below of  = userAccount, xshift = -2.5cm, it]     {Token.Deposit()};
	\node (tokenWithdraw)     [withdraw, below of  = userAccount, xshift = 2.5cm, it]     {Token.Withdraw()};
	\node (contractAccount)     [start, below of  = tokenWithdraw, xshift = -2.5cm]     {Contract Account};
	\node (rBalance)            [process, below of  = contractAccount, xshift = 2.5cm]     {Recipient Contract Balance};
	\node (sBalance)            [process, below of  = contractAccount, xshift = -2.5cm]     {Sender Contract Balance};
	\node (channelLoad)     [deposit, below of  = sBalance, xshift = -2.5cm, it]     {Channel.Create()\\Channel.Load()};
	\node (channelUnload)     [withdraw, below of  = sBalance, xshift = 2.5cm, it]     {Channel.Unload()};
	\node (channelCollect)     [withdraw, below of  = sBalance, xshift = 8cm, it]     {Channel.CollectPayment()};
	\node (channel)               [start, below of = channelUnload]  {Channel};
	\node (channelExtend)               [function, below of = channel, xshift = 2.5cm, it]  {Channel.ExpireTime()};
	\node (channelAbort)               [function, below of = channel, xshift = -2.5cm, it]  {Channel.Abort()};
	% Specification of lines between nodes specified above
	% with aditional nodes for description 
	
	\draw[->]             (userAccount) -- (tokenDeposit);
	\draw[<-]             (userAccount) -- (tokenWithdraw);
	\draw[<-]             (contractAccount) -- (tokenDeposit);
	\draw[->]             (contractAccount) -- (tokenWithdraw);
	\draw[<->]           (contractAccount) -- (rBalance);
	\draw[<->]           (contractAccount) -- (sBalance);
	\draw[<-]             (channelLoad) -- (sBalance);
	\draw[->]             (channelUnload) -- (sBalance);
	\draw[->]             (channelCollect) -- (rBalance);
	\draw[->]             (channelLoad) -- (channel);
	\draw[<-]             (channelUnload) -- (channel);
	\draw[<-]             (channelCollect) -- (channel);
	\draw[-]               (channelExtend) -- (channel);
	\draw[-]               (channelAbort) -- (channel);
	%{deposit}(contractAccount);
	%\draw[<-]             (userAccount) -- node[yshift=-0.5cm, text width=1cm]
	%{withdraw}(contractAccount);
	%\draw[<->]     (contractAccount) -- (channel);
	\end{tikzpicture}
	\caption[position=bottom]{Flow diagram of the unidirectional Payment channel protocol implementation in V Systems. Arrows represent the possible flow of funds. Each $Channel.function()$ call changes the state of the channel. 
	\label{fig:flowxLumiImpl}}
\end{figure}

Figure \ref{fig:flowxLumiImpl} shows the flow of funds between the xLumi channel's user, the contract account and its underlying channel. $Token.function()$ represents a call to the token contract, and $Channel.function()$ represents a call to the channel contract. Calling any functions within the channel contract changes the state of its underlying channel.

\subsection{Details of Implementation}
xLumi's V Systems Smart Contract will record the amount already collected from the contract, and future signed offline transactions must give an amount larger than this. This feature allows the recipient to collect funds to their own account and settle that transaction on the blockchain without having to close the payment channel. The recipient executes the smart contract $Channel.CollectPayment()$ function, with the signed payment message as input which contains the payment amount information, and $\left(amount - State.collected\right)$ number of VSYS Coins or VSYS tokens will be transferred to the recipient's contract balance. The state value $State.collected$ will then be updated to the newest amount. This differs from other implementations of payment channels that settle the state and distribute funds only once the payment channel is closed. The recipient is allowed to take funds out of the xLumi channel whenever they wish, without closing the payment channel

This also means that the payer cannot unload their funds until the xLumi channel is closed, but the channel will allow them to store increasingly large amounts of funds. This ensures that the recipient knows at least how many tokens are contained within the channel, and the payer cannot cheat by signing an offline transaction, then withdrawing all the funds from the channel. However, since the payer is the owner of the channel, they can close the payment channel whenever they wish, once closed, there is a \textbf{grace period} of several days in which the funds will still be locked, and the recipient is still allowed to collect any remaining signed transactions they have yet to broadcast. After this grace period,  funds remaining in the contract can be unloaded by the owner/payer of the channel.

The recipient should continuously monitor the status of the channel, in case the payer decides to abort the channel, the recipient should collect any uncollected funds within the \textbf{grace period}, therefore, the recipient should confirm the state of the contract at least once every \textbf{grace period}. Although, should they wish to continue using the channel without having to open a new one, the payer is allowed to extend the expiration time of the channel, and load  more funds into it.

\subsection{Example of xLumi Contract Use}
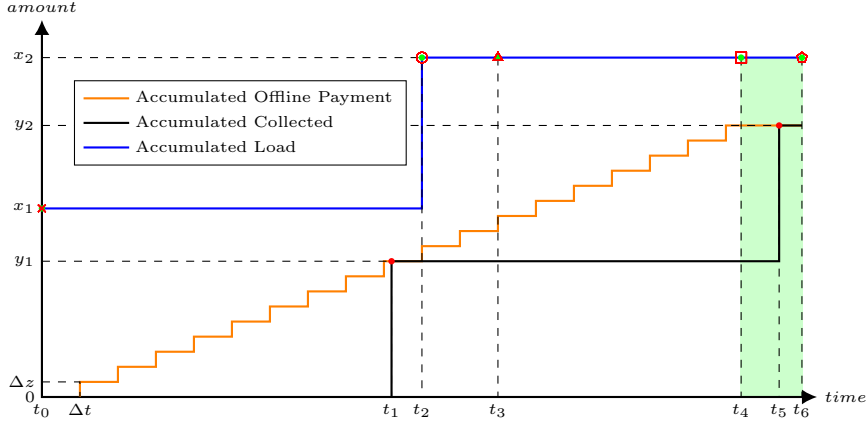
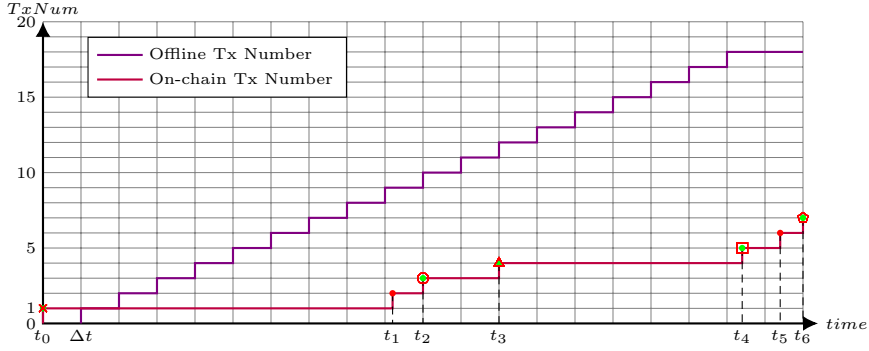
\begin{figure}[H]
	
	\centering
	\tiny
	\begin{subfigure}{0.72\textwidth}
		\begin{tikzpicture}[scale=1]
		% Draw axes
		
		\draw[draw = green!20,double=green!20, double distance=0.8cm] (9.6,0) -- (9.6,4.5);
		
		\draw [<->,thick] (0,5) node (yaxis) [above] {$amount$}
		|- (10.2,0) node (xaxis) [right] {$time$};

		% Draw X value step function
		\draw [blue, thick] (0,2.5) -| (5,4.5) -- (10,4.5);
		
		% Draw Z value step function
		\draw [orange, thick] (0.5,0) |- (1,0.2) |- (1.5,0.4)  |- (2,0.6) |- (2.5,0.8) |- (3,1) |- (3.5,1.2)  |- (4,1.4) |- (4.5,1.6) |- (5,1.8) |- 	(5.5,2)  |- (6,2.2) |- (6.5,2.4) |- (7,2.6) |- (7.5,2.8)  |- (8,3) |- (8.5,3.2) |- (9,3.4) |- (10,3.6);
		
		% Draw Y value step function
		\draw [black, thick] (4.6,0) |- (9.7,1.8) |- (10,3.6);
		
		% Calculate the intersection of the lines a_1 -- a_2 and b_1 -- b_2
		% and store the coordinate in c.
		\coordinate (d_1) at (4.6, 1.8);
		\coordinate (d_2) at (9.7, 3.6);
		\coordinate (b_1) at (0, 2.5);
		\coordinate (b_2) at (5, 4.5);
		\coordinate (b_3) at (6, 4.5);
		\coordinate (b_4) at (9.2, 4.5);
		\coordinate (b_5) at (10, 4.5);
		\coordinate (c_1) at (0.5, 0.2);
		% Draw lines indicating intersection with y and x axis. Here we use
		% the perpendicular coordinate system
		\draw[dashed] (yaxis |- c_1) node[left] {$\Delta z$}
		-| (xaxis -| c_1) node[below] {$\Delta t$};
		\draw[dashed] (yaxis |- d_1) node[left] {$y_1$}
		-| (xaxis -| d_1) node[below] {$t_1$};
		\draw[dashed] (yaxis |- b_1) node[left] {$x_1$}
		-| (xaxis -| b_1) node[below] {$t_0$};
		\draw[dashed] (yaxis |- b_2) node[left] {$x_2$}
		-| (xaxis -| b_2) node[below] {$t_2$};
		\draw[dashed] (6,4.5) -- (6,0) node[below] {$t_3$};
		\draw[dashed] (9.2,4.5) -- (9.2,0) node[below] {$t_4$};
		\draw[dashed] (yaxis |- d_2) node[left] {$y_2$}
		-| (xaxis -| d_2) node[below] {$t_5$};
		\draw[dashed] (10,4.5) -- (10,0) node[below] {$t_6$};
		\draw(0,0) node[left] {$0$};

		%		\draw[dashed] (yaxis |- d_3) node[left] {$v_3$}
		%		-| (xaxis -| d_3) node[below] {$t_3$};
		%		
		%		\draw[dashed] (yaxis |- d_4) node[left] {$m$}
		%		-| (xaxis -| d_4) node[below] {$t_4$};
		%		% Draw a dot to indicate intersection point
		\fill[red] (d_1) circle (1.2pt);
		\fill[red] (d_2) circle (1.2pt);
		\fill[green] (b_1) circle (1.2pt);
		\fill[green] (b_2) circle (1.2pt);
		\fill[green] (b_3) circle (1.2pt);
		\fill[green] (b_4) circle (1.2pt);
		\fill[green] (b_5) circle (1.2pt);
		
		\draw[red] plot[mark = x] (b_1);
		\draw[red] plot[mark = o] (b_2);
		\draw[red] plot[mark = triangle] (b_3);
		\draw[red] plot[mark = square] (b_4);
		\draw[red] plot[mark = pentagon] (b_5);
		\begin{customlegend}[legend cell align=left, %<= to align cells
		legend entries={ % <= in the following there are the entries
			Accumulated Offline Payment,
			Accumulated Collected,
			Accumulated Load
		},
		legend style={at={(4.8,4.2)},font=\tiny}] % <= to define position and font legend
		% the following are the "images" and numbers in the legend
		\addlegendimage{draw = orange, thick}
		\addlegendimage{draw = black, thick}
		\addlegendimage{draw = blue, thick}
		\end{customlegend}
		\end{tikzpicture}
		\caption{The changes in contract states through a number of user interactions.}
	\end{subfigure}
	
	\begin{subfigure}{0.72\textwidth}
		\begin{tikzpicture}[scale=1]
		% Draw axes
		
		\draw[gray,very thin] (0,0) grid [xstep = 0.5, ystep = 0.2)] (10,4);
		
		\draw [<->,thick] (0,4) node (yaxis) [above] {$Tx Num$}
		|- (10.2,0) node (xaxis) [right] {$time$};
		
		% Draw Z value step function
		\draw [violet, thick] (0.5,0) |- (1,0.2) |- (1.5,0.4)  |- (2,0.6) |- (2.5,0.8) |- (3,1) |- (3.5,1.2)  |- (4,1.4) |- (4.5,1.6) |- (5,1.8) |- (5.5,2)  |- (6,2.2) |- (6.5,2.4) |- (7,2.6) |- (7.5,2.8)  |- (8,3) |- (8.5,3.2) |- (9,3.4) |- (10,3.6);
		
		% Draw Y value step function
		\draw [purple, thick] (0,0) |- (4.6,0.2) |- (5,0.4) |- (6,0.6) |- (9.2, 0.8) |- (9.7, 1.0) |- (10, 1.2)-- (10, 1.4); 
		
		% Calculate the intersection of the lines a_1 -- a_2 and b_1 -- b_2
		% and store the coordinate in c.
		\coordinate (p_1) at (0, .2);
		\coordinate (p_2) at (4.6, .4);
		\coordinate (p_3) at (5, .6);
		\coordinate (p_4) at (6, .8);
		\coordinate (p_5) at (9.2, 1);
		\coordinate (p_6) at (9.7, 1.2);
		\coordinate (p_7) at (10, 1.4);
		% Draw lines indicating intersection with y and x axis. Here we use
		% the perpendicular coordinate system
		
		\draw (0,0) node[below] {$t_0$};
		\draw (0.5,0) node[below] {$\Delta t$};
		\draw[densely dashed] (4.6,0.4) -- (4.6,0) node[below] {$t_1$};
		%	\draw (4.6,0) node[below] {$t_1$};
		\draw[densely dashed] (5,0.6) -- (5,0) node[below] {$t_2$};
		%	\draw (5,0) node[below] {$t_2$};
		\draw[densely dashed] (6,0.8) -- (6,0) node[below] {$t_3$};
		%	\draw (6,0) node[below] {$t_3$};
		\draw[densely dashed] (9.2,1) -- (9.2,0) node[below] {$t_4$};
		%	\draw (9.2,0) node[below] {$t_4$};
		\draw[densely dashed] (9.7,1.2) -- (9.7,0) node[below] {$t_5$};
		%	\draw (9.7,0) node[below] {$t_5$};
		\draw[densely dashed] (10,1.4) -- (10,0) node[below] {$t_6$};
		%	\draw (10,0) node[below] {$t_6$};
		
		\draw (0,0) node[left] {$0$};
		\draw (0,0.2) node[left] {$1$};
		\draw (0,1) node[left] {$5$};
		\draw (0,2) node[left] {$10$};
		\draw (0,3) node[left] {$15$};
		\draw (0,4) node[left] {$20$};

		\fill[green] (p_1) circle (1.2pt);
		\fill[red] (p_2) circle (1.2pt);
		\fill[green] (p_3) circle (1.2pt);
		\fill[green] (p_4) circle (1.2pt);
		\fill[green] (p_5) circle (1.2pt);
		\fill[red] (p_6) circle (1.2pt);
		\fill[green] (p_7) circle (1.2pt);

		\draw[red] plot[mark = x] (p_1);
		\draw[red] plot[mark = o] (p_3);
		\draw[red] plot[mark = triangle] (p_4);
		\draw[red] plot[mark = square] (p_5);
		\draw[red] plot[mark = pentagon] (p_7);
		\begin{customlegend}[legend cell align=left, %<= to align cells
		legend entries={ % <= in the following there are the entries
			Offline Tx Number,
			On-chain Tx Number
		},
		legend style={at={(4,3.8)},font=\tiny}] % <= to define position and font legend
		% the following are the "images" and numbers in the legend
		\addlegendimage{draw = violet, thick}
		\addlegendimage{draw = purple, thick}
		\end{customlegend}
		\end{tikzpicture}
		\caption{The difference between total transactions and the number of transactions recorded on the blockchain.}
	\end{subfigure}
	\caption{An example of the evolution of states of an xLumi contract through time. The green dots represent on-chain transactions by the payer. The red dots represent on-chain transactions by the recipient. The green area is the grace period after the payer aborts.}
	\label{fig:implementationFig}
\end{figure}

In this section we will be giving a graphical example of an xLumi contract. Figure \ref{fig:implementationFig} mimics an xLumi contract being used, clearly showing online and offline transactions.

Payer creates the channel at $t_0$ with $x_1$ loaded amount. From $\Delta t$, payer pays $\Delta z$ to recipient every $\Delta t$. At $t_1$, recipient collects $y_1$ from the channel. After that, the Payer loads $x_2 - x_1$ to the channel at $t_2$ and changes the accumulated load to $x_2$. Payer then extends the expiration time at $t_3$. Payer aborts the channel at $t_4$ and updates the expiration time to $t_6$. Recipient observes the the payer broadcasting an \textbf{abort} transaction and collects the last $y_2 - y_1$ amount from the channel at $t_5$. The final amount recipient collected from the channel is $y_2$. Finally the Payer unloads $x_2 - y_2$ from the channel. At time $t_6$ the total number of transaction fees saved is 11.

	\section{Extensions}
	\label{sec:ext}
	The design decisions made for the V Systems implementation of xLumi were driven by two main factors: ease of use, and security of funds. It is posed as an open question to the reader if there are ways to add useful features to the payment channel without compromising these two factors.
	
	\subsection{Bidirectional Payment Channel}
	The simplest method of creating a two-way payment channel using this protocol is to open two unidirectional payment channels in both directions. If we truly wanted to only utilise one payment channel, the implementation must change significantly.
	
	One possible way to turn this into a two-way payment channel protocol is by taking the idea from how some implementations of the lightning Network intends to use the Eltoo update \cite{decker2018eltoo} to the Bitcoin protocol. The basic idea is to allow newer transactions to overwrite the state of payment channel, this way, even if a malicious party attempts to broadcast an older transaction, the other party can simply overwrite the state by broadcasting a newer transaction. This can be done by adding another index value that is only allowed to increase, if a new transaction to update the state of the payment channel is broadcast, it will only overwrite the state if it has a higher index value than the previous transaction.
	
	If this were to be done, several features of xLumi would need to be changed. Firstly, since this would be a two-way payment channel, it would become impossible for either party to take funds out before the payment channel closes. Secondly, in a bidirectional payment channel, the rules to closing the channel must be different, xLumi allows only the payer to close the channel at any point, and giving a period in which the state can still be updated. A bidirectional implementation may need to allow both parties to close the channel. It will still maintain the advantage of ensuring the correct state without relying on punishments or third parties.
	
	\subsection{Grace Period}
	
	There is also a choice in the decision of the period in which the recipient is still allowed to broadcast an update to the state after the closing of the channel. Since the payer is allowed to close the channel at any point in this implementation, it is necessary for the recipient to be allowed some time to still update the state before the funds are distributed accordingly, in case there are still uncollected signatures. This grace period can either be decided within the implementation to ensure fairness between the two parties, or allow the payer to decide the allowed period. Allowing the payer to select the allowed period has the advantage of creating a more flexible payment channel, and allows the security for the recipient to be changed. The decision should be agreed on upon by both parties before the opening of the payment channel, otherwise the recipient can simply disregard the channel and refuse to provide the product or services. This potentially adds an extra level of complexity to opening payment channels.
	
	\subsection{Payment Channel Network}
	Just like the Lightning Network, nodes can be connected through unidirectional payment channels to allow payments to anyone within the connected network. An extra addition to the protocol may be required to allow for hashed time-locked contracts whereby parties can only spend the funds if they solve some hash function or the transaction times out from the time lock. This is not a trivial addition to the protocol, and any such implementations should undergo significant scrutiny in their design to ensure security.
	
	\newpage
	
	Section \ref{sec:uses} describes the expected uses of this payment channel, while there are many other potential applications, we discuss the most obvious areas of uses, and further clarify our design decisions. Section \ref{Summary} concludes this paper, and reiterates the general problems that payment channels face, with how xLumi attempts deals with these issues.
	
	\section{Usage Cases}
	\label{sec:uses}
	Figure \ref{fig:implementationFig} demonstrates a payment channel where all of its functionalities are utilised. The total number of saved transactions depend on the frequency of off-chain transactions compared to on-chain transactions. We define three levels of usage for xLumi contracts, each with their minimum number of on-chain transactions for the payer.
	
	\begin{itemize}
		\item Level one: An xLumi channel that utilises only the \textbf{create} function. In this case, only one on-chain transactions is required.
		\item Level two: An xLumi channel that utilises the \textbf{create}, \textbf{unload}, and \textbf{abort} functions. In this case, at least three on-chain transactions are required.
		\item Level three: An xLumi that utilises all possible functionalities, \textbf{create}, \textbf{unload}, \textbf{extend expiration time}, \textbf{load}, and \textbf{abort} functions. In this case, the minimum number of on-chain transactions depend on the number of times the expiration time needs to be extended, and the number of times the channel needs to be loaded.
	\end{itemize}
	
	It is likely that a level two usage of the contract is suitable for most use cases of payment channels. Therefore, users should consider utilising a payment channel if they expect to pay the recipient more than three times.
	
	\subsection{Micropayments}
	Payment channels allow users to send very small amounts of funds, since there is no need to pay for transaction fees, it would be feasible to send payments much lower than the transaction fee. This may be the case for service users with their providers. Services that charge by use may be impossible with traditional blockchain transactions, since paying transaction fees can become very costly compared to small usage amounts. Payment channels open up the possibility for service providers to charge by use, and they can simply collect the funds once it accumulates high enough. xLumi will allow service providers to collect the funds at any point without the need to close the payment channel, so there is minimal opportunity cost to using the channel.
	
	For transactions where the transferred amount far exceeds the fees, it should be more practical to simply use traditional method of payment on the blockchain.
	
	\subsection{High Frequency Transactions}
	Payment channels allow instantaneous payments without needing transaction fees, so they could potentially be used for extremely high frequency transactions. Since blockchains have a limited transaction throughput, large volumes of transactions may need to span several blocks and therefore have a delay for settlement. For example, exchanges need to settle a very high volume of trades at all times, and will usually record the transactions and broadcast a final state to the relevant blockchains. Payment channels can replicate this function.
	
	It is expected that these will be the main scenarios for payment channel usage, high frequency and micropayments. These transactions will likely have a well defined payer and recipient and a unidirectional payment channel is sufficient. A notable usage case for bidirectional payment channels is its use in payment channel networks. However, it is very much possible to generate two unidirectional payment channels to achieve the same result as proposed in \cite{decker2015fast}.
	
	\subsection{Data Storage}
	Decentralised data storage may rely on splitting data into a number of small parts and sending them to storage nodes \cite{benet2018filecoin, benet2014ipfs}. This means that data objects will not be sent and stored in its entirety directly. Payment channels can enable such decentralised storage methods by allowing data centres to charge per piece of data sent and successfully stored, reducing the risk of users overpaying for data storage.
	
	\section{Conclusion}
	\label{Summary}
	Payment channels can allow blockchains to scale without changes to protocol. Large amounts of transactions can be done out of the blockchain's consensus framework, saving on computing power and storage space. Despite this, payment channels have a series of challenges it must overcome.
	
	The main problem with payment channels is that it is very difficult to enforce the final state of the payment channels, since every signed transaction is a valid on-chain transaction, it is very difficult to ensure that no party cheats by broadcasting an older state that benefits them.
	
	\subsection{Monitoring Malicious Activity}
	
	In order to prevent cheating parties from broadcasting a state that benefits them and stealing funds, it is crucial that there is some mechanism to prevent this from happening. The Lightning Network \cite{poon2016bitcoin} uses something called the punishment payment channel. The party that observes a dishonest broadcasting of an older state can take all of the funds in the channel within a certain period of time. This requires both parties to continuously monitor the blockchain to ensure neither party cheats. This can be somewhat expensive if the implemented time period is too short (since both parties must check the state at least once every period). However, if the time period is set to be too long, whoever decides to close the channel will have to wait for that amount of time before the funds are safely in their wallet, unless the channel is closed cooperatively.
	
	While xLumi retains the need to monitor the payment channel, this responsibility only lies with the recipient. It is also not possible for either party to act in a malicious manner, because it is not possible to broadcast an older state than the one currently recorded on chain.
	
	\subsection{Data Loss}
	It is also possible that either party simply loses a few transactions from corruption or software bugs. In certain payment channel implementations, that party may end up being unable to close the channel and obtain any funds.
	
	This may be possible to solve by utilising a trusted third party to store transactions and signatures for particular payment channels. For most payment channel implementations, it should be impossible for the third party to use the signatures to obtain the funds. However, this may cause privacy concerns, and introduce an extra cost for using payment channels.
	
	The risk of data loss can be mitigated in xLumi since the state of the channel can be updated without the need to close the channel. This risk only applies to the recipient, who is the only party required to store signatures. The recipient can update the state of the channel whenever they wish, if they deem the amount paid to be sufficiently large such that the risk of losing it to data loss exceeds the amount saved in transaction fees. It is also possible to broadcast older states, if only newer states were lost.
	
	\subsection{Summary of the Unidirectional Payment Channel Protocol}
	A payment channel protocol called Super Luminal ("xLumi") has been proposed in this paper that deals with the difficulty of determining the final state of the channel by controlling the funds using smart contracts in a very simple manner. The contracts store several important values including the amount deposited into the payment channel, the amount already taken from the channel by the recipient and an expiration timestamp of the channel. xLumi contracts only allow actions that increase these state variables, ensuring that only newer states can be recorded, it is therefore not possible for either party to cheat, this bypasses the need for any punishment system in place for cheating parties, or third parties to ensure the correctness of the payment channel state.
	
	xLumi also allows the recipient to obtain funds from the payment channel whenever they wish without closing the xLumi channel, which contrasts with current implementations of the Lightning Network which has to close the channel before funds are distributed. This allows the recipient to control the amount of risk they are willing to take. 
	
	The offline signatures by the payer cannot be taken back, but they can be lost to software bugs or hardware losses. Once the amount signed accumulates to some amount, the recipient can settle the signed payments on-chain, and the funds will be sent to their account and the risk of losing signatures can be mitigated.
		
	The payer in an xLumi channels is allowed to close the payment channel at any point, which activates a time period in which the recipient can still settle payment signatures on-chain. This time period should be implemented such that it is not expensive for the recipient to scan the blockchain once per period.
	
	This payment channel protocol drastically reduces the number of interactions and complexity of opening a payment channel, and does not require users store a new secret for every off-chain transaction made. The lower number of interactions reduces the number of places where the protocol can go wrong.
	
	The maximum amount of information needed to be stored locally while using an xLumi payment channel is a single signature. Once a newer signed payment message is received, the recipient can delete any old signatures. Without the need for a large number of secrets to be stored, mistakes in storing the keys is much less likely, and the memory required to maintain the channel does not increase with its age.
	
	\subsection{Final Remarks}
		
	Blockchains are often seen as a method of taking trust out of industries, with the majority of community valuing trustless protocols, because of this, blockchain protocols must prove that they cannot be broken except with negligible probability, and that any attacks would be infeasible. However, fundamentally, proofs are done in theory, and implementations may end up deviating from the theoretical model, it is necessary for users of a protocol to have some level of trust that the protocol is robust.
	
	It may be that blockchains should not be seen as a method of taking trust out of existing industries, but because it creates a much lower baseline level of trust required for interactions than conventional methods, it can serve as a stepping stone for forming deeper relationships. The philosophy behind our design, is intended to lower the barrier of trust required to interact with one another, allowing relationships to be built where it was not possible before.

	\section{Acknowledgment}
	We thank everyone on the V Systems team for assistance with testing, revising and development, who worked tirelessly to ensure the quality of the protocol. We are also immensely grateful to Alex Yang, for their comments on an earlier version of the manuscript, although any errors are our own and should not tarnish the reputations of these esteemed persons.
	
	\bibliographystyle{plain}
	\bibliography{yningc}
	
	% if need to cite the ref, first add the articles in yningc.bib, then you can cite it, it will automatically sort the refs.
\end{document}